\documentclass[aps,twocolumn,prl,showpacs,superscriptaddress]{revtex4-1}
\usepackage{graphicx}
\usepackage{amsmath}
\usepackage{amssymb}
\usepackage{color}
\usepackage{wasysym}
\usepackage{enumerate}
\usepackage{ifthen}

\newcommand \be{\begin{equation}}
\newcommand \ee{\end{equation}}

\newcommand \jconf{\left\{ J_{ij} \right\}}

\newboolean{mywordcount}
\setboolean{mywordcount}{false}

\begin{document}

\ifthenelse {\not \boolean{mywordcount}}
{
\title{
Numerical construction of the Aizenman-Wehr metastate
}

\author{A. Billoire}
\affiliation{Institute de Physique Th\'eorique, CEA Saclay and CNRS,
  91191 Gif-sur-Yvette, France}
\author{L.A. Fernandez}
\affiliation{Departamento de F\'isica Te\'orica I, Universidad Complutense, 28040 Madrid, Spain}
\affiliation{Instituto de Biocomputaci\'on y F\'isica de Sistemas Complejos (BIFI), 50009 Zaragoza, Spain}
\author{A. Maiorano}
\affiliation{Dipartimento di Fisica, Universit\`a di Roma La Sapienza, I-00185 Rome, Italy}
\affiliation{Instituto de Biocomputaci\'on y F\'isica de Sistemas Complejos (BIFI), 50009 Zaragoza, Spain}
\author{E. Marinari}
\affiliation{Dipartimento di Fisica, Universit\`a di Roma La Sapienza, I-00185 Rome, Italy}
\affiliation{Nanotec, Consiglio Nazionale delle Ricerche, I-00185 Rome, Italy}
\affiliation{Istituto Nazionale di Fisica Nucleare, Sezione di Roma I, I-00185 Rome, Italy}
\author{V. Martin-Mayor}
\affiliation{Departamento de F\'isica Te\'orica I, Universidad Complutense, 28040 Madrid, Spain}
\affiliation{Instituto de Biocomputaci\'on y F\'isica de Sistemas Complejos (BIFI), 50009 Zaragoza, Spain}
\author{J. Moreno-Gordo}
\affiliation{Departamento de F\'isica Te\'orica I, Universidad Complutense, 28040 Madrid, Spain}
\affiliation{Instituto de Biocomputaci\'on y F\'isica de Sistemas Complejos (BIFI), 50009 Zaragoza, Spain}
\author{G. Parisi}
\affiliation{Dipartimento di Fisica, Universit\`a di Roma La Sapienza, I-00185 Rome, Italy}
\affiliation{Nanotec, Consiglio Nazionale delle Ricerche, I-00185 Rome, Italy}
\affiliation{Istituto Nazionale di Fisica Nucleare, Sezione di Roma I, I-00185 Rome, Italy}
\author{F. Ricci-Tersenghi}
\affiliation{Dipartimento di Fisica, Universit\`a di Roma La Sapienza, I-00185 Rome, Italy}
\affiliation{Nanotec, Consiglio Nazionale delle Ricerche, I-00185 Rome, Italy}
\affiliation{Istituto Nazionale di Fisica Nucleare, Sezione di Roma I, I-00185 Rome, Italy}
\author{J.J. Ruiz-Lorenzo}
\affiliation{Departamento de F\'isica and Instituto de Computaci\'on Cient\'ifica
  Avanzada (ICCAEx), Universidad de Extremadura, 06071 Badajoz, Spain}
\affiliation{Instituto de Biocomputaci\'on y F\'isica de Sistemas Complejos (BIFI), 50009 Zaragoza, Spain}
\date{\today}

\begin{abstract}
Chaotic size dependence makes it extremely difficult to take the
thermodynamic limit in disordered systems. Instead, the metastate,
which is a distribution over thermodynamic states, might have a smooth
limit. So far, studies of the metastate have been mostly mathematical.
We present a numerical construction of the metastate for the $d=3$
Ising spin glass. We work in equilibrium, below the critical
temperature. Leveraging recent rigorous results, our numerical
analysis gives evidence for a {\em dispersed} metastate, supported on
many thermodynamic states.
\end{abstract}


\maketitle
}{}

\emph{Introduction}. 
Symmetry breaking and phase transitions are well defined for systems of infinite size, $L=\infty$. Indeed, $L=\infty$ states
can be defined rigorously as Dobrushin-Lanford-Ruelle (DLR)
states~\cite{ruelle:04}. However, experiments are conducted
on systems of very large, but finite size. Therefore, one would like to interpret DLR states
as the $L\to\infty$ limit of states defined on a sequence of systems of growing size.
Indeed, for simple systems, e.g.\ ferromagnets, we know how to approach DLR states with finite-$L$ states  
by using suitable boundary conditions. Yet, for disordered
systems~\cite{parisi:94,young:98}, the connection between DLR and
finite-$L$ states is still much of a mistery~\cite{AW-MS:90}, most
particularly for
spin-glasses~\cite{NS,NS-proc:03,NS-book:13,Arguin:11,Read:14}.

\begin{figure}
\begin{center}
\includegraphics[width=1.0\columnwidth]{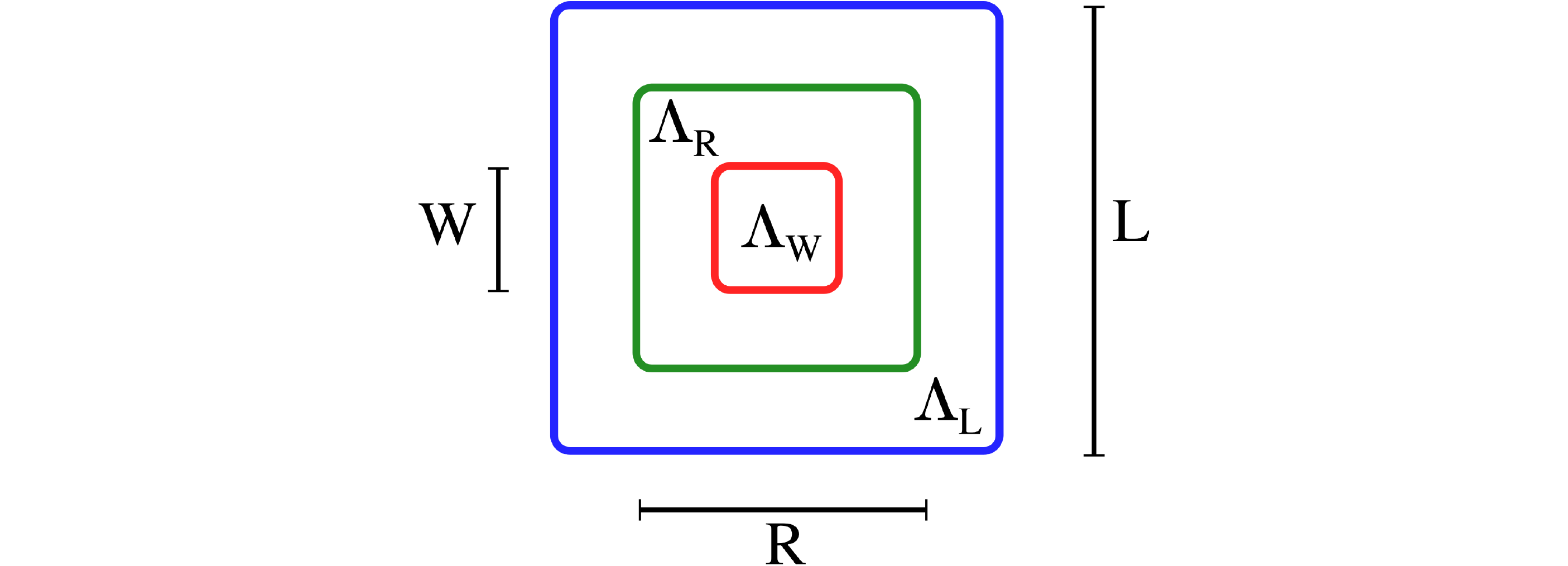}
\caption{Sketch of the AW metastate construction.}
\label{fig:AW}
\end{center}
\end{figure}

For the sake of concreteness, let us consider the standard model for
spin-glasses, the Edwars-Anderson model~\cite{EA:75} in spatial dimension
$d$. Ising spins $s_i=\pm1$, located in a size $L$ cube,
$\Lambda_L\subset\mathbb{Z}^d$ (see Fig.~\ref{fig:AW}), interact through a
nearest neighbor, bond disordered and strongly frustrated Hamiltonian:
\ifthenelse {\not \boolean{mywordcount}} {
\begin{align}
\label{eq:HEA}
H_{\mathcal{J},L}(\underline{s}) =-\sum_{\langle i,j\rangle}J_{ij}\,s_i\,s_j\;.
\end{align}
}{}
The quenched couplings $J_{ij}$ are independent and
identically distributed random variables ($J_{ij}=\pm1$ with $50\%$
probability, in our case).  We call $\mathcal{J}\equiv\jconf$ a disorder
sample. The finite-$L$ Gibbs state $\Gamma_{\mathcal{J},L}(\underline{s})=
\exp(-H_{\mathcal{J},L}(\underline{s})/T)/Z_{\mathcal{J},L}$ is a random state, as it
depends on the set of random couplings $\mathcal{J}$.

The problem in taking the large $L$ limit for spin-glasses defined by Eq.~\eqref{eq:HEA} is
caused by their chaotic size-dependence. Take a fixed, arbitrary yet finite
region (e.g. the \emph{measuring window} $\Lambda_W$ in Fig.~\ref{fig:AW}). The Gibbs
measure over $\Lambda_W$ changes chaotically when the system grows by the addition of
new couplings at the boundaries, while keeping previous couplings
unaltered. This extreme sensibility to changes at the boundaries motivated the
invention of the \emph{metastate} \cite{NS}, a probability distribution over
states with a (hopefully) smoother $L\to\infty$ limit.

The two main metastate definitions are Aizenman-Wehr's \cite{AW-MS:90} and
Newman-Stein's \cite{NS}. We shall focus on the former (the two definitions are
conjectured to be equivalent~\cite{NS-proc:03}, but the AW's is easier to
implement numerically).
The lattice $\Lambda_L$ in Fig.~\ref{fig:AW} is divided into an inner region
$\Lambda_R$, a cube of linear size $R$, and an outer region.  Consequently we
call internal couplings the set $\mathcal{I} \equiv \{J_{ij}
|i,j\in\Lambda_R\}$ and outer couplings $\mathcal{O} =
\mathcal{J}\setminus\mathcal{I}$. We proceed by (i) restricting our attention
to the measuring window $\Lambda_W$ of linear size $W$~\cite{Read:14}, (ii)
taking the average over the outer couplings, with \emph{fixed} internal
couplings, and (iii) sending to infinity all three length scales while
respecting $W \ll R \ll L$. If the limit exists (which is yet to be proven),
it is independent of the arbitrary choice for the fixed internal
couplings~\footnote{For the sake of completeness, let us briefly recall the
  construction of the Newman-Stein metastate. In this approach, the set of
  couplings $\mathcal{J}$ over the \emph{infinite} lattice $\mathbb{Z}^d$ is
  fixed from the outset. A sequence of growing finite regions $\Lambda_L$ is
  considered, see Fig.~\ref{fig:AW} (all the cubes $\Lambda_L$ are centered at
  the origin of $\mathbb{Z}^d$). The Hamiltonians in Eq.~\eqref{eq:HEA} are
  truncated to the cube $\Lambda_L$ by a choice of boundary contitions
  (e.g. free boundary conditions). We consider the Gibbs state,
  $\Gamma_{\mathcal{J},L}$ as restricted to the measuring window $\Lambda_W$.
  The Newman and Stein metastate\cite{NS} records the frequency by which each
  state appears while the system size grows. The convergence of this sequence
  of states is yet to be proven. However, at least when the size $W$ of the
  measuring window gets large, the resulting metastate is expected to be
  independent of the initial choice of couplings $\mathcal{J}$ (provide that
  this $\mathcal{J}$ are typical).}.

Yet, even if at a considerably lesser level of mathematical rigor, we must
recall that there has been some progress in the study of spin glasses. We have
numerical~\cite{Palassini:99,Ballesteros:00} and
experimental~\cite{Gunnarson:91} evidences for the existence of a phase
transition to a spin glass phase at low temperatures in $d=3$, at least in the
absence of an external magnetic field. We also have two major theoretical
frameworks that are applied to interpret experiments and simulations: the
replica symmetry breaking (RSB) theory \cite{MPV,reviewJSP} and the droplet
model \cite{BM-Scaling1,BM-Scaling2,FH-DM}. Which (if any) of these two
theories captures the nature of the spin glass phase in $d=3$ is being
debated \cite{Yucesoy-Billoire:12}.

In fact, a recent mathematical tour de
force~\cite{guerra,talagrand,panchenko} has shown that the RSB theory
provides the exact solution to the $d=\infty$ Sherrington-Kirkpatrick
model \cite{sherrington:75}. The common lore expects RSB to be also
valid above and at the upper critical dimension $d_\mathrm{U}=6$.  RSB
theory extends as well to $d<d_\mathrm{U}$ features found in the mean
field solution~\cite{Parisi-RSB}: many states (infinitely many in the
$L\to\infty$ limit), hierarchically organized, contribute to the Gibbs
measure, each one with a weight that depends on the disorder
realization. Consequences include the existence of de Almeida-Thouless
line~\cite{dealmeida:78} (the spin-glass phase transition survives in
the presence of a small external magnetic field \cite{PNAS4d}), or the
strong sample-to-sample fluctuations induced by the
non-self-averageness of several measurable quantities \cite{Janus:11}
(these observations \cite{PNAS4d,Janus:11} were, however, obtained by
simulating systems of finite size).

The alternative Droplet Model provides a much simpler scenario for the
spin-glass phase, where the Gibbs measure is a mixture of two spin-flip related
pure states. It follows that the spin-glass phase transition should disappear
when a magnetic field is applied [the field breaks the global spin-flip
  symmetry $H(\underline {s})=H(-\underline{s})$ in
  Eq.~\eqref{eq:HEA}].

The most recent mathematical analysis, based on metastates, has critically assessed
both the RSB theory and the Droplet Model.  Currently, we have three
mathematically consistent pictures for the spin-glass phase.  First, the
Droplet Model metastate is concentrated on a single trivial state (let us call
`trivial' a state which is a mixture of two pure states related by the global
spin-flip symmetry).  Second, we have the Chaotic Pairs
picture~\cite{{NS,NS-book:13}}, predicting a \emph{disperse} metastate (there
is a large number of states to choose from), yet each state is
trivial.  This non-trivial metastate is due to chaotic size-dependence:
by gradually increasing $L$, one obtains vastly different states.  Finally,
the RSB-metastate~\cite{Read:14} is disperse and every state drawn from it
contains the Parisi hierarchical tree of pure states. Alternatives to these
three pictures are much limited by recent rigorous results~\cite{Arguin:15}.

Read argues~\cite{Read:14} that one can partially discriminate between
these competing pictures for the metastate by studying the decay of a
particular correlation function averaged over the metastate,
$C_\rho(x)\sim 1/|x|^{d-\zeta}$ for large distances $|x|$, see
Eq.~\eqref{E_CX}. An exponent value $\zeta<d$ implies a disperse
metastate, thus ruling out the Droplet Model's metastate.
Furthermore, in the context of the RSB-metastate, the number of pure
states that can be resolved by studying a region of size $W$ is
exponentially large in $W^{d-\zeta}$.

To the best of our knowledge there has been only one numerical attempt to
study the metastate, by means of a non-equilibrium simulation
\cite{Young-MS}. Yet, the main debated points regard the equilibrium
metastate.  In fact, the only related issue addressed numerically by
equilibrium simulations has been non self-averageness~\cite{Marinari:98,
  Janus:11, Yucesoy-Billoire:12, Middleton:13, Billoire:14}.

Here we show that a numerical construction of the Aizenman and Wehr metastate
for the Edwards-Anderson model in $d=3$ is possible in present-day
computers. Our construction makes precise several hints by
Read~\cite{Read:14}. In particular, recall Fig.~\ref{fig:AW}, we show how big
the ratios of lengthscales $L/R$ , $R/W$ need to be to uncover metastate
properties. We also study the dependence on the fixed internal couplings, a
crucial issue that hast not yet been addressed quantitatively. We make
quantitative computations of overlap distributions and correlation functions
averaged over the AW metastate, thus computing the crucial $\zeta$
exponent. We find a value definitively smaller than $d=3$, which rules out the
Droplet Model metastate and leaves the Chaotic Pairs and the RSB metastates as
the remaining contenders.

\emph{Metastate averages and the Metastate-averaged state.} In the context
depicted by Fig.~\ref{fig:AW}, we consider model~\eqref{eq:HEA} endowed with
periodic boundary contitions (which makes irrelevant the location of
$\Lambda_R$ in $\Lambda_L$).
Let us consider the probability distribution of $\Gamma_{\mathcal{J},L}$ at fixed internal disorder $\mathcal{I}$, while sending $L\to\infty$ and averaging over the outer disorder $\mathcal{O}$:
\[
\kappa_{\mathcal{I},R}(\Gamma) = \lim_{L\to\infty} \mathbb{E}_\mathcal{O}\Big[\delta^{(F)}\left(\Gamma - \Gamma_{\mathcal{J},L}\right)\Big]
\]
If the limit $\kappa(\Gamma) = \lim_{R\to\infty}
\kappa_{\mathcal{I},R}(\Gamma)$ exists, it does not longer depend on the
internal disorder $\mathcal{I}$ and provides the AW metastate. The purpose of
the ``measuring window'' $\Lambda_W$ in Fig.~\ref{fig:AW} is avoiding boundary
effects, that may appear as long as $R$ is finite. Any measure is taken only
inside $\Lambda_W$, while bonds are fixed in $\Lambda_R$ in the metastate
definition.

We have two kind of averages, thermal averages over the Gibbs state $\langle
\cdots \rangle_\Gamma$ and averages over the metastate $[\cdots]_\kappa$, that
can be combined in different ways.  For example, the \emph{metastate averaged
  state} (MAS) $\rho(\underline{s})$ is defined via the average $\langle
\cdots \rangle_\rho \equiv [\langle \cdots \rangle_\Gamma]_\kappa$.

As seen from the measuring window $\Lambda_W$, a state $\Gamma(\underline{s})$
is a set of probabilities $\{p_\alpha\}_{\alpha=1,\ldots,2^{W^d}}$ over the
spin configurations in $\Lambda_W$.  In other words, it is a point on the
hyperplane defined by the equation $\sum_\alpha p_\alpha=1$, $p_\alpha\geq
0$. In this sense, the metastate is a probability distribution over this
hyperplane. The MAS $\rho(\underline{s})$ is the average of this distribution,
and it is itself a point on the hyperplane (hence, the MAS is a state itself).

\emph{The numerical construction of the metastate.}
We simulate the EA model ($8\le L \le 24$) sampling
spin configurations at equilibrium by
a combination of Metropolis single spin flip Monte Carlo and Parallel
Tempering~\cite{PT}. All the data shown are measured at the
lowest simulated temperature $T=0.698\simeq 0.64 T_c$, well below the
critical temperature $T_c=1.102(3)$ \cite{JANUS-crit:13}.
Equilibration was assessed on a
sample-by-sample basis~\cite{JANUS:10} and, for the largest systems,
it required the use of multi-site multi-spin coding
(MUSI-MSC)~\cite{fernandez:15} (see~\cite{billoire:17} for details).

We repeat the computation for $\mathcal{N}_\mathcal{I}=10$ different internal
couplings $\mathcal{I}$ samples (indexed by $0\le \mathtt{i} < \mathcal{N}_\mathcal{I}$)
and, for each of these, we use $\mathcal{N}_\mathcal{O}=1\,280$
different outer disorder $\mathcal{O}$ realizations (indexed by $0\le \mathtt{o} < \mathcal{N}_\mathcal{O}$)
Thus we have a total of $\mathcal{N}_\mathcal{J}=12\,800$ samples and, for each sample $\mathcal{J}=\mathcal{I}\cup\mathcal{O}$, we simulate $m=4$ distinct real replicas.

We take $\mathcal{N}_\mathcal{I}\ll \mathcal{N}_\mathcal{O}$ because we expect
all inner disorder samples to be ``typical'' \cite{Read:14} when computing
metastate averages at $R \gg 1$. We found however sizable sample to
sample fluctuations for the system sizes we consider.

The average over the Gibbs state $\langle\cdots\rangle_\Gamma$ is estimated
via Monte Carlo thermal averages $\langle\cdots\rangle$ at fixed disorder
$\mathcal{J}$, i.e.\ for given indices $\mathtt{i}$ and $\mathtt{o}$.  The
average over the metastate is given by $[\cdots]_\kappa = \sum_\mathtt{o}
(\cdots) / \mathcal{N_O}$, and the one over the internal disorder by
$\overline{(\cdots)} = \sum_\mathtt{i} (\cdots) / \mathcal{N_I}$.  For
example, the MAS spin correlation function is given by
\begin{multline}
C_\rho(x) =\overline{\left[\langle s_0 s_x \rangle_\Gamma \right]_\kappa^2} =
\frac{1}{\mathcal{N_I}} \sum_\mathtt{i} \left( \frac{1}{\mathcal{N_O}} \sum_\mathtt{o} \langle s_0^\mathtt{i;o} s_x^\mathtt{i;o} \rangle \right)^2 =\\
= \frac{1}{\mathcal{N_I}} \sum_\mathtt{i} \frac{1}{\mathcal{N}_\mathcal{O}^{\,2}} \sum_\mathtt{o,o'} \langle s_0^\mathtt{i;o} s_x^\mathtt{i;o} s_0^\mathtt{i;o'} s_x^\mathtt{i;o'} \rangle
\sim |x|^{-\left(d-\zeta\right)}\;,
\label{E_CX}
\end{multline}
defining the Read's $\zeta$ exponent for $|x| \gg 1$.

We measure in $\Lambda_W$ the overlaps between any two real replicas --- let us call it $\underline{\sigma}$ and $\underline{\tau}$ --- sharing the same internal disorder (indexed by $\mathtt{i}$) and having external couplings indexed by $\mathtt{o}$ and $\mathtt{o'}$
\begin{equation}
  q_\mathtt{i;o,o'} \equiv \frac{1}{W^3}\sum_{x\in\Lambda_W} \sigma_x^\mathtt{i;o} \tau_x^\mathtt{i;o'}\;.
  \label{eq:qOOIrr}
\end{equation}
Actually, for each $\{\mathtt{i;o,o'}\}$, we have $m(m-1)/2$ contributions from different pairs of real replicas if $\mathtt{o}=\mathtt{o'}$ and $m^2$ otherwise.

The main objects of our numerical study are the probability density functions (pdf) of the overlaps:
\begin{eqnarray*}
P(q) = \frac{\sum_\mathtt{i} P_\mathtt{i}(q)}{\mathcal{N_I}}\; &,&\quad P_\mathtt{i}(q) = \frac{1}{\mathcal{N_O}} \sum_\mathtt{o} \langle \delta(q-q_\mathtt{i;o,o}) \rangle,\\
P_\rho(q) = \frac{\sum_\mathtt{i} P_{\rho,\mathtt{i}}(q)}{\mathcal{N_I}}\; &,& \quad P_{\rho,\mathtt{i}}(q) =\frac{1}{\mathcal{N}_\mathcal{O}^{\,2}} \sum_\mathtt{o,o'} \langle \delta(q-q_\mathtt{i;o,o'}) \rangle.
\end{eqnarray*}
While $P(q)$ is the usual pdf already measured in many numerical simulation of spin glasses, $P_\rho(q)$ is the pdf of the overlap over the MAS. Although $P_\rho(q) \to \delta(q)$ for $W\to\infty$ \cite{Read:14}, the scaling of its variance is informative
\begin{equation}
\chi_\rho = \sum_{x\in\Lambda_W} C_\rho(x) =  W^d \int q^2 P_\rho(q)\,dq \sim W^\zeta\;.
\label{eq:chiA}
\end{equation}


\begin{figure}
\begin{center}
\includegraphics[width=\columnwidth]{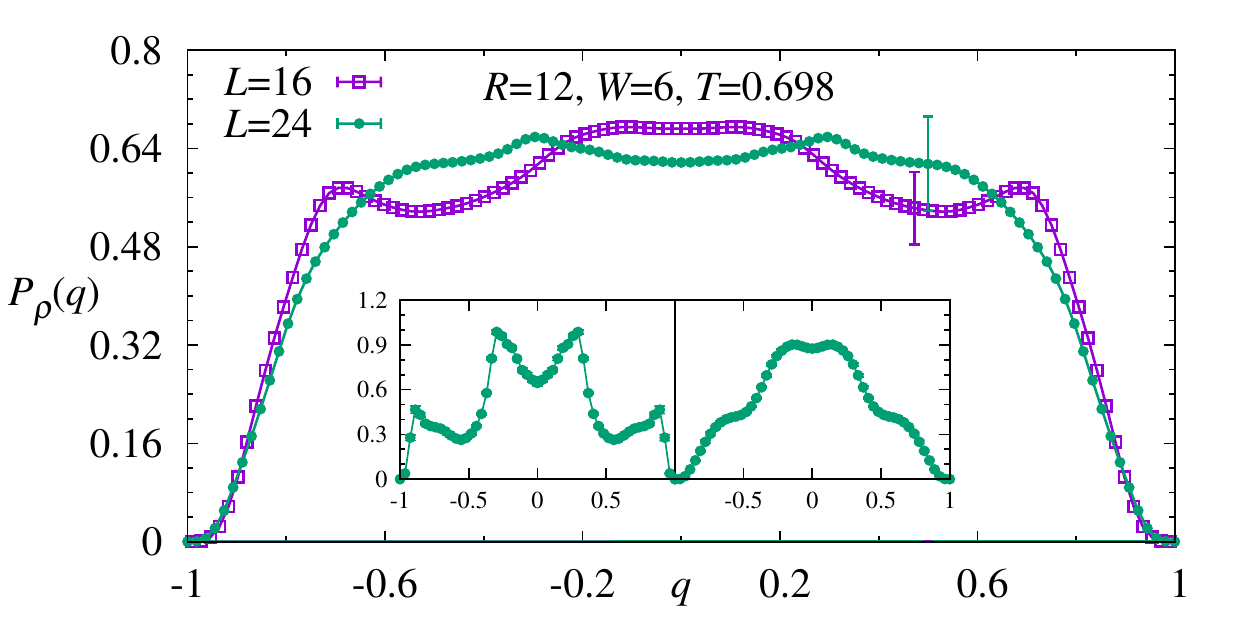}
\caption{Main plot: The MAS overlap distribution $P_\rho(q)$
  with $R=12$ at $T=0.698 \simeq 0.64 T_c$ shows no statistical significant dependence
  on the lattice size $L$ (the error bars, computed from fluctuations on $\mathcal{I}$,
  are shown for one value of $q$ only for sake of clarity).
  Insets: $P_{\rho,\mathtt{i}}(q)$ for two specific configurations of the inner disorder
  $\mathcal{I}$ (the error bars, computed from fluctuations on $\mathcal{O}$, are smaller
  than the data points).}
\label{fig:PA}
\end{center}
\end{figure}

\begin{figure}
\begin{center}
\includegraphics[width=\columnwidth]{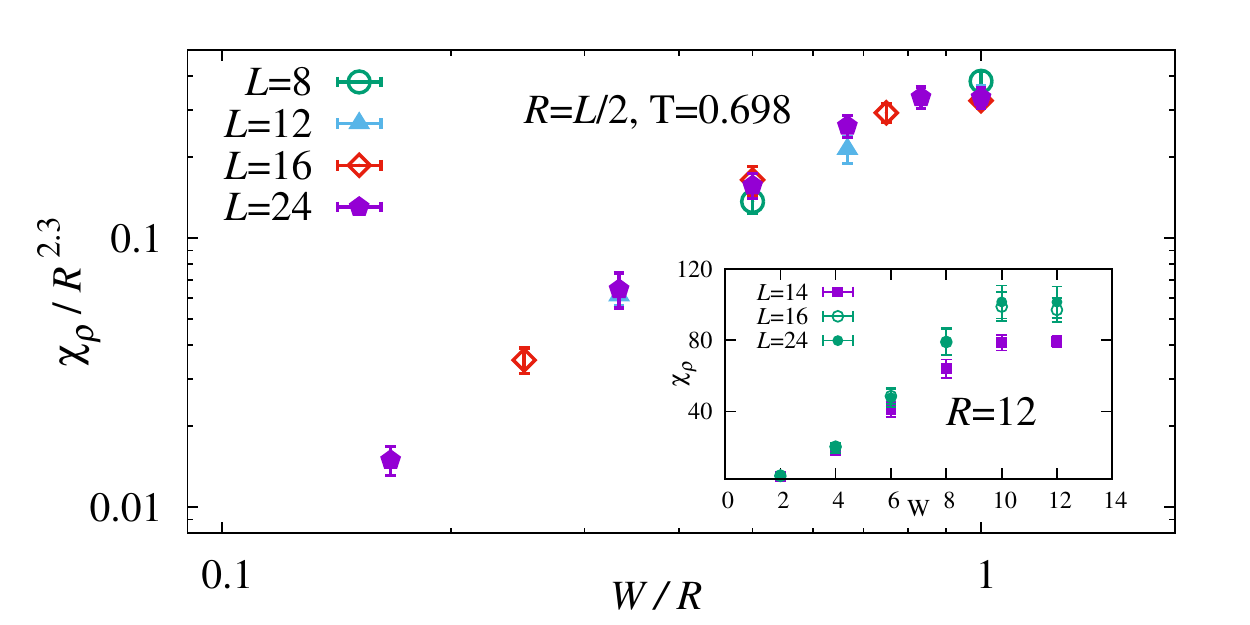}
\caption{Main plot: Collapse of MAS susceptibility data measured with $R=L/2$ at $T=0.698 \simeq 0.64 T_c$.
	Inset: Deviations from the asymptotic behavior are evident only for $R/L>3/4$.}
\label{fig:scaling}
\end{center}
\end{figure}

\emph{Numerical results}.  Taking the limit of large $L$ in simulations
actually amounts to asking how small the ratios $R/L$ and $W/R$ need to be, in
order to find results which are \emph{safe} (to a given accuracy).

In Fig.~\ref{fig:PA} (main plot) we see that that the MAS $P_\rho(q)$
measured with $R=12$ and both $L=24$ and $L=16$ are statistically compatible,
suggesting that $R/L=3/4$ is already a safe choice.
The error bars are rather large, because the dependence of $P_{\rho,\mathtt{i}}(q)$
on the internal disorder sample is unexpectedly strong for the values
of $W$ and $R$ we are using (as shown by the insets in Fig.~\ref{fig:PA}).

A similar check can be performed with the MAS susceptibility $\chi_\rho$
(see Fig.~\ref{fig:scaling}).
In the inset we see that, fixing $R=12$, all data with $R/L\le3/4$ are statistical
compatible, while data with $R/L=6/7$ show significant deviations even for
small $W$ values. Hereafter we safely fix $R=L/2$.

The main panel of Fig.~\ref{fig:scaling} shows data for the MAS susceptibility
$\chi_\rho$ measured with the safe ratio $R=L/2$ (which is statistically
equivalent to the limiting condition $R\ll L$) and different ratios $W/R$.
Data have been rescaled according to the following scaling law
\begin{equation}
\chi_\rho(W,R) = R^\zeta f(W/R) = W^\zeta g(W/R)\;,
\end{equation}
which is compatible with Eq.~(\ref{eq:chiA}) if
$f(x) \propto x^\zeta$ for $x\ll 1$ and $g(0)=\text{const.}$
First of all we note that the physical behavior we would like to measure
in the limit $W/R\ll 1$ actually extends up to $W/R \approx 0.75$, where
corrections to the asymptotic power law appear.
Fitting data with $W/R < 0.75$ we estimate Read's exponent $\zeta=2.3\pm 0.3$.

\begin{figure}
\begin{center}
\includegraphics[width=\columnwidth]{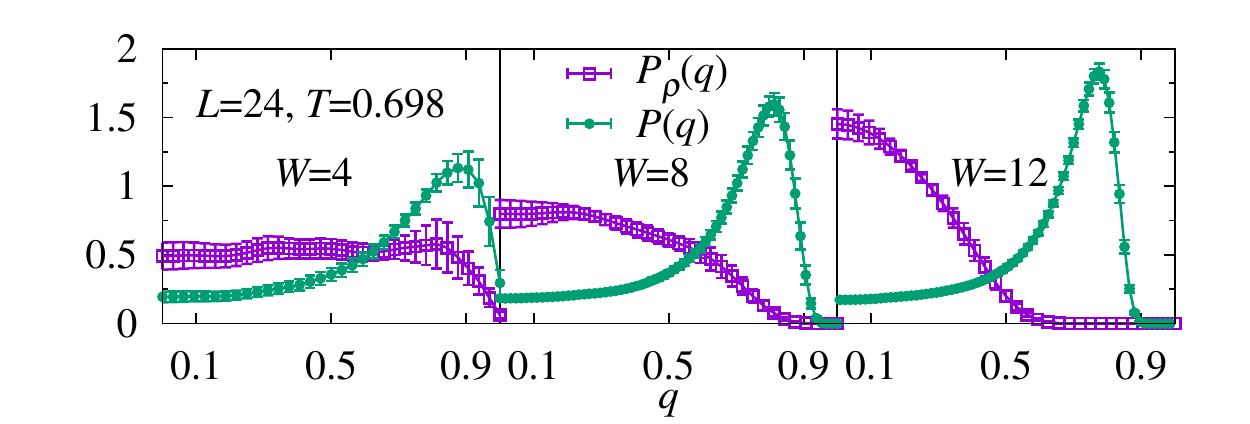}
\caption{$P_\rho(q)$ and $P(q)$ for $L=24$, $R=L/2$, $T=0.698$ and different values of the measuring window size, $W=4,8,12$.}
\label{fig:pqcomp}
\end{center}
\end{figure}

Finally we show in Fig.~\ref{fig:pqcomp} the size dependence of both
$P_\rho(q)$ and $P(q)$, for $L=24$ (the largest simulated), $R=L/2$
and varying $W$. For a dispersed metastate in the thermodynamic limit the two
distributions are different.


\emph{Discussion and conclusions.}
We have shown that state-of-the-art numerical simulations of spin glasses
in $d=3$ allow for the construction of the AW metastate.
Numerical data suggest that the limiting conditions $1\ll W \ll R \ll L$
can be relaxed to $W/R,R/L\approx 3/4$ without changing substantially
the thermodynamic physical behavior.
These are unexpected very good news.

From the numerical construction of the AW metastate we have obtained
quantitative information on the nature of the spin glass phase in $d=3$.
The metastate average overlap distribution $P(q)$ and the MAS $P_\rho(q)$ are
significantly distinct objects already at moderate sizes. We cannot
extrapolate safely to the thermodynamic limit, and sample to sample
fluctuations are still important at the accessible system
sizes. Nevertheless we have found a convincing scaling law for the MAS
susceptibility, and an estimate of $\zeta(d=3)=2.3(3)$, strongly suggesting
$\zeta< d$.

\begin{figure}
\begin{center}
\includegraphics[width=0.9\columnwidth]{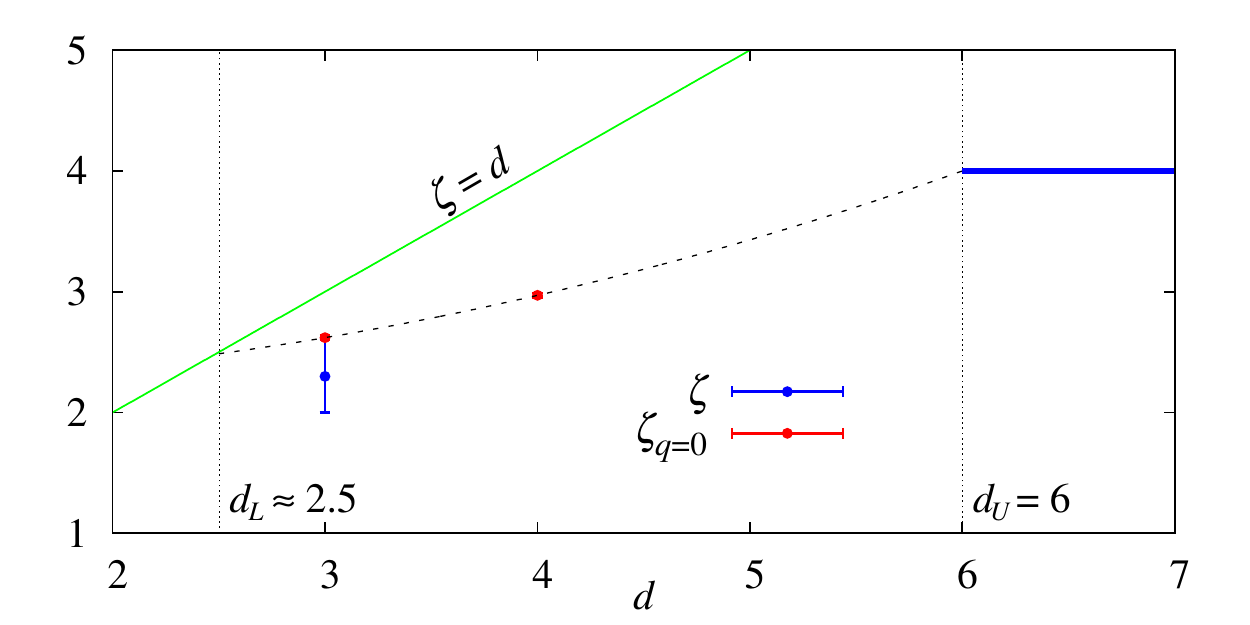}
\caption{The exponent $\zeta$ as a function of $d$.}
\label{fig:cartoon}
\end{center}
\end{figure}

Read's exponent $\zeta$ is related to the number of different states that can
be measured in a system of size $W$ as $\log n_\text{states}\sim W^{d-\zeta}$
\cite{Read:14}. Such a number diverges in the thermodynamic limit as long as
$\zeta<d$, supporting the picture of a metastate with infinitely many states.
In Fig.~\ref{fig:cartoon} we summarize our knowledge about the $\zeta$
exponent.  At and above the upper critical dimension $d_U=6$, where mean field
exponents are correct, $\zeta=4$~\cite{DD:99,RFSGBook}.  Assuming $\zeta(d)$
is a continuous and monotonically non-decreasing function, the inequality
$\zeta<d$ still holds slightly below $d_U$.  In the present work we find
$\zeta(d=3)=2.3(3)$ (blue point in Fig.~\ref{fig:cartoon}).  An alternative
estimate of the $\zeta$ exponent comes from the decay of the 4-spins spatial
correlation function conditional to the $q=0$ sector,
$C_4(x) = \overline{[\langle \sigma_0^\mathtt{i;o} \tau_0^\mathtt{i;o}\sigma_x^\mathtt{i;o} \tau_x^\mathtt{i;o}\rangle_{\Gamma|q_\mathtt{i;o,o}=0}]_\kappa} \sim |x|^{-(d-\zeta_{q=0})}$
for $|x|\gg 1$:
$\zeta_{q=0}(d=3) = 2.62(2)$~\cite{JANUS:09,JANUS:10,UNPUB} and
$\zeta_{q=0}(d=4) = 2.97(2)$~\cite{NICOLAO:14}. Read conjectured that
$\zeta_{q=0}=\zeta$~\cite{Read:14}. These
estimates are shown by red points in Fig.~\ref{fig:cartoon}~\footnote{The conjectured relation~\cite{RFSGBook} 
  $\zeta=(d+2-\eta_c)/2$, where $\eta_c$ is the anomalous dimension at the
  critical point, does not work in $d=3,4$~\cite{JANUS:09,JANUS:10,UNPUB,NICOLAO:14}.}.  A gentle
interpolation of the $\zeta$ estimates (dashed line in Fig.~\ref{fig:cartoon})
seems to meet the $\zeta=d$ condition very close to the current best estimate
for the lower critical dimension $d_L\approx 2.5$ \cite{Boettcher}.

In conclusions, all the numerical evidences strongly support the existence
of a spin glass metastate dispersed over infinitely many states for $d=3$
(and probably down to the lower critical dimension).
These findings are incompatible with the Droplet Model, while are compatible
with both the Chaotic Pais picture and the Replica Symmetry Breaking scenario.

\ifthenelse {\not \boolean{mywordcount}}
{
\section{Acknowledgments}
This project has received funding from the European Research Council
(ERC) under the European Union's Horizon 2020 research and innovation
program (grant agreement No 694925). We were partially supported by
MINECO (Spain) through Grant Nos. FIS2012-35719-C02-01,
FIS2013-42840-P, FIS2015-65078-C2, FIS2016-76359-P (contract partially
funded by FEDER), and by the Junta de Extremadura (Spain) through
Grant No. GRU10158 (partially funded by FEDER). Our simulations were
carried out at the BIFI supercomputing center (using the
\emph{Memento} and \emph{Cierzo} clusters), at the TGCC supercomputing
center in Bruy\`eres-le-Ch\^atel (using the \emph{Curie} computer,
under the allocation 2015-056870 made by GENCI) and at ICCAEx
supercomputer center in Badajoz (\emph{GrInFishpc} and
\emph{ICCAExhpc}). We thank the staff at BIFI, TGCC and ICCAEx
supercomputing centers for their assistance.

}{}

\end{document}